\documentclass[aps, prb, twocolumn, showpacs, groupedaddress, showkeys]{revtex4}
\usepackage[dvips]{color}
\usepackage{amsfonts}
\usepackage{amsmath}
\usepackage{amssymb}
\usepackage{graphicx}
\usepackage{mathrsfs}
\usepackage{bbm}
\usepackage{bm}
\usepackage{multirow}
\usepackage[sort&compress]{natbib}
\usepackage{dcolumn}%

\begin{document}

\newcommand{\dd}{d}
\newcommand{\pd}{\partial}
\newcommand{\myU}{\mathcal{U}}
\newcommand{\myr}{q}
\newcommand{\Urho}{U_{\rho}}
\newcommand{\myalpha}{\alpha_*}
\newcommand{\bd}[1]{\mathbf{#1}}
\newcommand{\Eq}[1]{Eq.~(\ref{#1})}
\newcommand{\Eqn}[1]{Eq.~(\ref{#1})}
\newcommand{\Eqns}[1]{Eqns.~(\ref{#1})}
\newcommand{\Figref}[1]{Fig.~\ref{#1}}
\newtheorem{theorem}{Theorem}
\newcommand{\me}{\textrm{m}_{\textrm{e}}}
\newcommand{\sgn}{\textrm{sign}}
\newcommand*{\bfrac}[2]{\genfrac{\lbrace}{\rbrace}{0pt}{}{#1}{#2}}
\newcommand{\figsize}{7.8cm}

\preprint{DOS}

\title{Density of States for Warped Energy Bands}

\author{Nicholas A. Mecholsky}
 \email{nmech@vsl.cua.edu}
 \thanks{Corresponding Author}

\author{Lorenzo Resca}
 \email{resca@cua.edu}
 \homepage{http://physics.cua.edu/people/faculty/resca.cfm}

\author{Ian L. Pegg}
 \email{ianp@vsl.cua.edu}

\affiliation{
Department of Physics and Vitreous State Laboratory\\
The Catholic University of America\\
Washington, DC 20064}

\author{Marco Fornari}
 \email{marco.fornari@cmich.edu}
 \homepage{http://www.phy.cmich.edu/people/fornari/}

\affiliation{
Department of Physics\\
Central Michigan University\\
Mount Pleasant, Michigan 48858}

\date{7/13/15}

\begin{abstract}
An angular effective mass formalism previously introduced is used to study the density of states in warped and non-warped energy bands. Band warping may or may not increase the density-of-states effective mass. Band ``corrugation," referring to energy dispersions that deviate ``more severely" from being twice-differentiable at isolated critical points, may also vary independently of density-of-states effective masses and band warping parameters. We demonstrate these effects and the superiority of an angular effective mass treatment for valence band energy dispersions in cubic materials. We also provide some two-dimensional physical and mathematical examples that may be relevant to studies of band warping in heterostructures and surfaces. These examples may also be useful in clarifying the interplay between possible band warping and band non-parabolicity for non-degenerate conduction band minima in thermoelectric materials of corresponding interest.
\end{abstract}

\pacs{72.10.Bg, 02.30.Mv, 72.20.Pa, 71.18.+y, 72.80.Cw}
\keywords{band warping, Density of States}

\maketitle


\section{Introduction} 
Effective mass approximations have been central to analyze and understand band structures of materials near critical points in the Brillouin zone (BZ) and their major physical consequences.  However, some basic formulae of that formalism have been misused for energy band dispersions that are not twice-differentiable. This has been further confused with band non-parabolicity effects in Taylor expansions. In a previous paper,\cite{MRPF} a more rigorous theory for dealing with a broad class of energy band dispersions has been developed. Here we advance that theory by applying its formalism to properly determine the density of states (DOS) and the DOS effective mass. The development of this formalism is necessary for energy dispersions at critical points that are ``warped,'' hence, conventional formulae involving second-order differentials are invalid.  

We begin by reviewing in Sec.\ \ref{sec:sec1} basic results of our previous treatment. In Sec.\ \ref{sec:DOS} we derive general expressions for the DOS and the DOS effective mass. We recover standard results for twice-differentiable ellipsoidal and hyperbolic energy dispersions in Sec.\ \ref{sec:easy}. Then we move on to study warped energy bands that are not twice-differentiable. In Sec.\ \ref{sec:warped} we apply our results to typical degenerate valence bands in cubic materials. Considerable differences emerge between our proper evaluations of the DOS effective masses and those reported in original papers.\cite{Dresselhaus, laxmavroides, mavroideslaxPR107in1957, Lawaetz1971} We discuss those differences in Sec.\ \ref{sec:Kittel} and Sec.\ \ref{sec:Lawaetzetal}. Subsequently, we focus on two-dimensional physical and mathematical models where the distinction between effects of band warping or ``corrugation'' and band non-parabolicity can be analytically demonstrated (Sec.\ \ref{sec:Two-Dimensional}). Some features of those examples may be useful in clarifying the interplay between possible band warping and band non-parabolicity in non-degenerate conduction bands of materials that exhibit remarkable thermoelectric properties.\cite{Chen2013, ParkerPRL2013, Singh2015May, HermanKortum1968, LentBowenDow1986, ValdiviaBarberis1995, Lach-habPapaconstantopoulos2002} Finally, we draw some conclusions and propose further inquiries in Sec.\ \ref{sec:Conclusions}.

\section{Angular Dependent Energy Dispersion}\label{sec:sec1}
We have previously considered\cite{MRPF} energy dispersions around a point $\bd{k}_0$ in a crystal BZ in the form of
\begin{align}\label{eqn:angularform}
E(k_r,\theta,\phi) &= E_0 + k_r a_1(\theta,\phi) \nonumber\\
&\phantom{=}+ k_r^2 a_2(\theta,\phi) + k_r^3 a_3(\theta,\phi) + \ldots.
\end{align}
Here, $k_r = |\bd{k} - \bd{k}_0|$ is the radial distance between a generic point at $\bd{k}$ in the BZ and the point of expansion at $\bd{k}_0$. The latter may be any point of special interest in the BZ, or a ``critical point,'' where the energy expansion has a null first-order differential.\cite{Bassani, GP} The polar angles $\theta$ and $\phi$ refer to the polar spherical coordinates of $\bd{k} - \bd{k}_0$.

It is essential to appreciate that \Eq{eqn:angularform} provides a much more general dispersion relation than commonly considered functions that admit multi-dimensional Taylor series expansions in Cartesian coordinates. That is so because \Eq{eqn:angularform} requires only the existence of a one-dimensional Taylor series expansion in each radial direction across $\bd{k}_0$. This is a much more limited requirement that can be reasonably expected of any physical band structure that allows one-dimensional transport of quasi-particles in any direction.\cite{MRPF} Besides ordinary quadratic bands, \Eq{eqn:angularform} thus includes ``warped bands," which are not twice-differentiable at isolated points, by definition. Typical examples of warped bands derive from original models of Dresselhaus \textit{et al.} and Kane.\cite{Dresselhaus, Kane1956}

Mathematically, band warping must be unambiguously distinguished from band non-parabolicity. The latter derives from higher-order terms $a_m(\theta, \phi)$ with $m>2$ in \Eq{eqn:angularform}. Conversely, band warping depends exclusively on the shape of the $a_2(\theta,\phi)$ term, which provides a dimensionless angular effective mass surface in Rydberg atomic units.\cite{MRPF} 

For an ordinary quadratic band, associated with a second-order differential and its curvature, $a_2(\theta,\phi)$ assumes the form
\begin{equation}\label{eqn:quadpolar}
a_2(\theta, \phi) = \frac{\sin ^2(\theta) \left(m_x \sin ^2(\phi )+m_y \cos^2(\phi )\right)}{m_x m_y \me^{-1}}+\frac{\cos^2(\theta)}{m_z \me^{-1}} 
\end{equation}
in a coordinate system of principal axes, with diagonal effective masses $m_x, m_y, m_z$, while $m_e$ is the ordinary electron mass. Any other form of $a_2(\theta, \phi)$ corresponds to a warped band, which cannot be exclusively described in terms of diagonal effective masses.

One may formally derive expressions for the DOS corresponding to the general energy expansion in \Eq{eqn:angularform}. In this paper, we focus on explicit DOS expressions for band warping, although we generalize our considerations at least to one type of band non-parabolicity, namely that of an overall energy dispersion of the form $E(k_r,\theta,\phi) = R(k_r) f(\theta,\phi)$, where $R(k_r)$ is a monotonically increasing function and all $a_i (\theta, \phi) = f(\theta, \phi)$. In this paper, we do not further consider any linear term in the energy expansion, thus assuming a null first-order differential at a ``critical point.''\cite{MRPF, Bassani, GP}

\section{Calculations of DOS}\label{sec:DOS}
In a crystal, the single-band DOS at energy $E$, within $\dd E$, is basically defined as
\begin{equation}\label{eqn:DOSdef}
g(E) = g_s \frac{V}{(2 \pi)^3} \int  \delta(E(\bd{k}) - E) \, \dd^3 \bd{k},
\end{equation}
where $g_s$ is a possible spin degeneracy, $V$ is the volume of the direct-lattice primitive cell, and $E(\bd{k})$ represents a single energy band in the BZ over which the $\dd^3 \bd{k}$ integration is performed. The integral may be evaluated after performing a transformation to ($E$, $\theta$, $\phi$) coordinates. The delta function can thus be handled by reducing the integral to the surface having given energy $E$ inside the BZ.\cite{Bassani, GP, Ashcroft}

\subsection{The DOS of Warped and Non-Warped Bands}\label{sec:dos}
Approaching a critical point $\bd{k}_0$ in the BZ, let us ignore band non-parabolicity effects for the moment and consider an energy dispersion (without any linear term) in the form\cite{MRPF}
\begin{equation}\label{eqn:f}
E(k_r, \theta, \phi) = E_0 + \frac{\hbar^2 k_r^2}{2 \me} f(\theta,\phi).
\end{equation}
In \Eq{eqn:f} we imply a partition of the unit $(\theta,\phi)$ sphere $S^2$ in a region $\mathcal{R}_+$ where $f(\theta, \phi) > 0$ and a region $\mathcal{R}_-$ where $f(\theta, \phi) < 0$, so that $\mathcal{R}_+ \cup \mathcal{R}_- = S^2$. This definition of $f(\theta, \phi)$ must refer to a single band, which may or may not be degenerate with other bands at $\bd{k}_0$. Typically, though not necessarily,\cite{Chen2013, ParkerPRL2013, Singh2015May} non-degenerate bands at $\bd{k}_0$ are not warped, corresponding to analytic maxima, minima, or saddle points. Conversely, degenerate bands are commonly warped.\cite{MRPF, Dresselhaus, Kane1956} 

For optical transitions, the joint density of states (JDOS) can be similarly considered.\cite{Bassani, GP} Both conduction and valence bands can be expressed as individual terms having the form of \Eq{eqn:f}. For the JDOS we may then define a joint $F(\theta, \phi)$ as the sum of the corresponding two (absorbing and emitting) $f$-contributions. The same formalism that we develop in this paper for the DOS thus essentially applies to the JDOS as well.  

In order to proceed with the integrations in \Eq{eqn:DOSdef}, we may first scale the Cartesian coordinates by letting $k_i' = \frac{\hbar k_i}{\sqrt{2 \me}}$. The energy dispersion thus becomes
\begin{equation}\label{eqn:Edef}
E(k_r, \theta, \phi) = E_0 + k_r'^2 f(\theta,\phi).
\end{equation}
We may then introduce a new variable $E'= E(\bd{k}') - E_0$, so that
\begin{equation}\label{eqn:DOSbasic}
g(E) = g_s \frac{V}{(2 \pi)^3} \left( \frac{2 \me}{\hbar^2} \right)^{3/2} \int  \delta(E' - (E - E_0)) \, \dd^3 \bd{k}'.
\end{equation}
In polar coordinates we have
\begin{align}\label{eqn:transformations}
k_x' &= \sqrt{\frac{E'}{f(\theta, \phi)}} \sin \theta \cos \phi, \nonumber\\
k_y' &= \sqrt{\frac{E'}{f(\theta, \phi)}} \sin \theta \sin \phi, \\
k_z' &= \sqrt{\frac{E'}{f(\theta, \phi)}} \cos \theta. \nonumber
\end{align}
Accordingly, we may perform a change of variables to spherical coordinates ($k_r'$, $\theta$, $\phi$), where $k_r'$ is defined implicitly through \Eq{eqn:Edef}. Notice that regions of positive $E'$ ($E > E_0$) correspond to $\mathcal{R}_+$ and regions of negative $E'$ ($E < E_0$) correspond to $\mathcal{R}_-$, so that all variables in \Eq{eqn:transformations} are real.

The Jacobian for this transformation is
\begin{equation}
J (E', \theta, \phi) = \sqrt{\frac{E'}{f(\theta, \phi)}} \frac{\sin \theta}{2 f(\theta,\phi)^2} > 0 , \quad \forall (\theta,\phi) \in S^2.
\end{equation}
With the change of variables to $(E', \theta, \phi)$, the DOS integral thus becomes 
\begin{widetext}
\begin{equation}\label{eqn:DOSformal}
g(E) = g_s \frac{V}{(2 \pi)^3} \left( \frac{2 \me}{\hbar^2} \right)^{3/2}   \int_{-\infty}^{\infty} \! \int_{0}^{\pi} \!\! \int_{0}^{2 \pi} \sqrt{\frac{E'}{f(\theta, \phi)}} \frac{\sin \theta}{2 f(\theta,\phi)^2}  \delta(E' - (E - E_0)) \, \dd \phi \dd \theta \dd E'.
\end{equation}
\end{widetext}

Now, if surface integrals over the unit sphere converge, we may split those integrals over regions of positive and negative $f(\theta,\phi)$, so that the energy integration immediately yields
\begin{equation}\label{eqn:DOS}
g(E) = \left\{ \begin{array}{ll} 
g_s \frac{V}{(2 \pi)^3} \left( \frac{2 \me}{\hbar^2} \right)^{3/2} \sqrt{E - E_0} \left[ C_{+} \right] , & E > E_0 ,\\
g_s \frac{V}{(2 \pi)^3} \left( \frac{2 \me}{\hbar^2} \right)^{3/2} \sqrt{E_0 - E} \left[ C_{-} \right] , & E < E_0 ,
\end{array} \right. 
\end{equation}
where we have defined
\begin{equation}\label{eqn:Cdefs}
C_{\pm} = \iint\limits_{\mathcal{R}_{\pm}} \frac{\sin \theta}{2 (\pm f(\theta,\phi))^{3/2}} \, \dd \theta \dd \phi.
\end{equation}

However, integration over the angular variables may not formally converge, as in the ideal case of a saddle point dispersion extending to infinity.\cite{Bassani} That is a theoretical extrapolation, however, because the BZ is actually finite, and so must be any band structure within it. Introduction of an energy-dependent cutoff parameter may thus be required, which should further take into account the onset of any significant band non-parabolicity. In any such case, the energy integration must be taken last, since $C_{\pm}$ also become functions of energy. However, the presence of the delta function can still make this last integration over energy relatively straightforward.  We provide an example of that in Sec.\ \ref{sec:hyperbolic}. 

\subsection{Generalization to Monotonically Non-Parabolic Bands}

We can readily extend the preceding formalism to energy dispersions of the form $E' = R(k_r') f(\theta,\phi)$, where $R(k'_r)$ is any monotonically increasing function of $k_r'$. The corresponding coordinate transformations then become 
\begin{align}\label{eqn:nonparabolic}
k_x' &= R^{-1}\left[\frac{E'}{f(\theta, \phi)}\right] \sin \theta \cos \phi, \nonumber\\
k_y' &= R^{-1}\left[\frac{E'}{f(\theta, \phi)}\right] \sin \theta \sin \phi, \\
k_z' &= R^{-1}\left[\frac{E'}{f(\theta, \phi)}\right] \cos \theta. \nonumber
\end{align}
The inverse function $R^{-1}$ of $R$ must exist and it has been introduced in \Eq{eqn:nonparabolic}. The DOS thus becomes
\begin{widetext}
\begin{equation}\label{eqn:DOSnonparabolic}
g(E) = g_s \frac{V}{(2 \pi)^3} \left( \frac{2 \me}{\hbar^2} \right)^{3/2} \int_{-\infty}^{\infty} \! \int_{0}^{\pi} \!\! \int_{0}^{2 \pi} R^{-1} \left[ \frac{E'}{f(\theta, \phi)}\right] \frac{\sin \theta}{2 f(\theta,\phi)^2}  \delta(E' - (E - E_0)) \, \dd \phi \dd \theta \dd E'.
\end{equation}
\end{widetext}
Equation (\ref {eqn:DOSformal}) may now be regarded as a special case of \Eq{eqn:DOSnonparabolic}, where $R^{-1}(x) = \sqrt{x}$. The energy integral may still be relatively straightforward to perform in \Eq{eqn:DOSnonparabolic} on account of the delta function.

\subsection{Two-dimensional DOS}
For further illustration, let us consider the two-dimensional evaluation of the DOS, according to the expression
\begin{equation}
g(E) = g_s \frac{V}{(2 \pi)^2} \int  \delta(E(\bd{k}) - E) \, \dd^2 \bd{k}.
\end{equation}
Close to a critical point $\bd{k}_0$ in the BZ, and ignoring band non-parabolicity, the two-dimensional energy dispersion becomes
\begin{equation}\label{basictwodimensional}
E(k_r, \theta) = E_0 + \frac{\hbar^2 k_r^2}{2 \me} f(\theta),
\end{equation}
where $f(\theta)$ is now a function of a single angular variable. The Jacobian of the transformation is simply
\begin{equation}
J_2 = \frac{1}{2 |f(\theta)|},
\end{equation}
and the DOS thus becomes
\begin{align}\label{eqn:2DDOS}
g(E) &= g_s \frac{V}{(2 \pi)^2} \left( \frac{2 \me }{ \hbar^2} \right) \times \nonumber\\ &\phantom{=} \int_{-\infty}^{\infty} \! \int_{0}^{2 \pi}  \frac{\delta(E' - (E - E_0))}{2 |f(\theta)|} \, \dd \theta \dd E'.
\end{align}
Again, we must integrate over regions of positive and negative $f(\theta)$ separately. Namely, the interval $(0,2 \pi)$ must be split into $\mathcal{R}_+$ and $\mathcal{R}_-$ regions, where $f(\theta)$ is either positive or negative, respectively.  Assuming that corresponding $\theta$-integrals converge, this yields
\begin{equation}\label{eqn:DOS2D}
g(E) = \left\{ \begin{array}{ll} 
g_s \frac{V}{(2 \pi)^2} \left( \frac{2 \me }{ \hbar^2} \right) \left[ C_{+} \right] , & E > E_0\\
g_s \frac{V}{(2 \pi)^2} \left( \frac{2 \me }{ \hbar^2} \right) \left[ C_{-} \right] , & E < E_0 ,
\end{array} \right.
\end{equation}
where 
\begin{equation}\label{eqn:Cdefs2D}
C_{\pm} = \int\limits_{\mathcal{R}_{\pm}} \frac{1}{2 |f(\theta)|} \, \dd \theta.
\end{equation}

\section{Twice-Differentiable Three-Dimensional Energy Dispersions}\label{sec:easy}
\subsection{Ellipsoidal Energy Dispersions}
Let us begin by considering the basic case of an ellipsoidal energy dispersion of the form
\begin{equation}\label{eqn:generalellipsoidal}
E(k_x, k_y, k_z) = \frac{\hbar^2}{2 m_x} k_x^2 + \frac{\hbar^2}{2 m_y} k_y^2 +  \frac{\hbar^2}{2 m_z} k_z^2 + E_0 ,
\end{equation}
where all three principal masses $m_x$, $m_y$, and $m_z$ have the same sign. In a four-dimensional space where $E(k_x, k_y, k_z)$ refers to the fourth dimension, that provides a paraboloid with a minimum (maximum) if all three masses are positive (negative). For positive masses, we may rescale the corresponding coordinates as $k'_i = \frac{\hbar k_i}{\sqrt{2 m_i}}$. The DOS is then expressed as
\begin{align}
g(E) &= g_s \frac{V}{(2 \pi)^3} \frac{2^{3/2} \left( m_x m_y m_z \right)^{1/2}}{\hbar^3} \times \nonumber\\ &\phantom{=}\int  \delta(E' - (E - E_0)) \, \dd^3 \bd{k}',
\end{align}
where $E' = k_r^{'2}$, and $f(\theta,\phi) = 1$. Integration is straightforward, yielding
\begin{equation}\label{eqn:Ccalculation}
C_+ = 2 \pi, \quad C_- = 0.
\end{equation}
We thus recover the well-known result
\begin{equation}\label{eqn:DOSmasses}
g(E) = \left\{ \begin{array}{ll} 
g_s \frac{V}{(2 \pi)^2} \frac{2^{3/2} \left( m_x m_y m_z \right)^{1/2}}{\hbar^3} \sqrt{E - E_0} ,  & E > E_0 , \\
0 , & E < E_0 .
\end{array} \right.
\end{equation}
In the special case where all masses equal a single effective mass, $m_x = m_y = m_z = m_*$, corresponding to a spherical energy dispersion, we have 
\begin{equation}\label{eqn:DOSeffectivemass}
g(E) = \left\{ \begin{array}{ll} 
g_s \frac{V}{(2 \pi)^2} \left( \frac{2 m_*}{\hbar^2} \right)^{3/2} \sqrt{E - E_0} , & E > E_0\\
0 , & E < E_0 .
\end{array} \right.
\end{equation}
Comparing Equations (\ref{eqn:DOSmasses}) and (\ref{eqn:DOSeffectivemass}), the DOS effective mass is ususally defined as $m_* = \left( m_x m_y m_z \right)^{1/3}$.

If all three masses are initially negative, essentially equivalent results can be obtained for the DOS by switching the signs of all three masses to positive, while correspondingly switching the signs of all energies to negative in the preceding equations, starting with \Eq{eqn:generalellipsoidal}.

\subsection{Hyperbolic Energy Dispersions}\label{sec:hyperbolic}
A hyperbolic energy dispersion still has the form of \Eq{eqn:generalellipsoidal}, but some of the masses have opposite signs. For the sake of simplicity, although without any major loss of generality, let us posit that $m_x = m_y = - m_z = m_e$ in \Eq{eqn:generalellipsoidal}. In that case, the angular effective mass surface $f(\theta,\phi)$ in \Eq{eqn:quadpolar} and \Eq{eqn:f}, and its corresponding $\mathcal{R_{\pm}}$ regions, are
\begin{align}
f(\theta, \phi) &= - \cos (2 \theta),\nonumber\\
\mathcal{R}_{+} &= \left\{ (\theta, \phi) \, \bigg| \, \frac{\pi}{4}< \theta < \frac{3 \pi}{4}\right\}, \\
\mathcal{R}_{-} &= \left\{ (\theta, \phi) \, \bigg| \, 0 < \theta < \frac{\pi}{4} \, \textrm{or} \, \frac{3 \pi}{4}< \theta < \pi \right\}.
\end{align}
As expected for an endless hyperbolic dispersion, the surface integrals do not converge and a `spherical' cutoff radius $R_c$ must be introduced.\cite{Bassani} In turn, this introduces an energy-angle relation at $R_c$, namely $\pm E' = - R_c^2 \cos 2(\frac{\pi}{4} \pm \epsilon)$, where $\epsilon$ represents the angular increment from $\frac{\pi}{4}$ corresponding to the intersection of the sphere of radius $R_c$ and the hyperboloid of constant $E'$. Now the surface integrals based on \Eq{eqn:Cdefs} can be formally performed, yielding
\begin{widetext}
\begin{subequations}
\begin{align}\label{eqn:hyperbolicCs}
C_{+}(E';R_c) &= 2 \cdot  2 \pi \int_{\frac{\pi}{4} + \epsilon}^{\frac{\pi}{2}} \frac{\sin \theta}{ 2 (- \cos 2 \theta)^{3/2}} \, \dd \theta \nonumber\\&= \frac{- \cos \theta}{2 \sqrt{ - \cos 2 \theta}} \Bigg|_{\frac{1}{2} \cos^{-1} \left(\frac{E'}{R_c^2}\right)}^{\frac{\pi}{2}} = \frac{2 \pi}{\sqrt{2}}\frac{R_c}{\sqrt{E'}} + \frac{\pi}{\sqrt{2}} \frac{\sqrt{E'}}{R_c} - \frac{\pi}{4 \sqrt{2}} \frac{(E')^{3/2}}{R_c^2} + \ldots \\
C_{-}(E';R_c) &= 2 \cdot 2 \pi \int_{0}^{\frac{\pi}{4} - \epsilon} \frac{\sin \theta}{ 2 (\cos 2 \theta)^{3/2}} \, \dd \theta \nonumber\\&= \frac{\cos \theta}{2 \sqrt{\cos 2 \theta}} \Bigg|_{0}^{\frac{1}{2} \cos^{-1} \left(\frac{-E'}{R_c^2}\right)} = \frac{2 \pi}{\sqrt{2}}\frac{R_c}{\sqrt{-E'}} - 2 \pi + \frac{\pi}{\sqrt{2}} \frac{\sqrt{-E'}}{R_c} - \frac{\pi}{4 \sqrt{2}} \frac{(-E')^{3/2}}{R_c^2} + \ldots.
\end{align}
\end{subequations}
\end{widetext}
We may thus perform in \Eq{eqn:DOSformal} the energy-dependent angular surface integrals over $\theta$ and $\phi$ of $C_{\pm}$ and complete the energy integration via the delta function, thus obtaining
\begin{widetext}
\begin{equation}\label{eqn:hyperbolicDOS}
g(E) = \left\{ \begin{array}{ll} 
g_s \frac{V}{(2 \pi)^3} \left( \frac{2 \me}{\hbar^2} \right)^{3/2} \left[ \frac{2 \pi}{\sqrt{2}} R_c + \frac{\pi}{\sqrt{2}} \frac{E - E_0}{R_c} - \frac{\pi}{4 \sqrt{2}} \frac{(E - E_0)^2}{R_c^2} + \ldots \right] , & E > E_0 ,\\
g_s \frac{V}{(2 \pi)^3} \left( \frac{2 \me}{\hbar^2} \right)^{3/2} \left[ \frac{2 \pi}{\sqrt{2}} R_c - 2 \pi \sqrt{E_0 - E} + \frac{\pi}{\sqrt{2}} \frac{E_0 - E}{R_c} - \frac{\pi}{4 \sqrt{2}} \frac{(E_0 - E)^2}{R_c^2} + \ldots \right] , & E < E_0 .
\end{array} \right.
\end{equation}
\end{widetext}
This agrees with the form of the DOS around an $M_1$ saddle point, as derived on p.~157 of Ref.~\onlinecite{Bassani}, for example.  

Beyond any such treatment, which is limited to twice-differentiable energy dispersions, we must now proceed to apply our angular effective mass formalism to more general calculations of the DOS in complex situations of warped energy bands.

\section{The DOS Effective Mass for Warped Energy Bands}\label{sec:warped}
\subsection{The DOS Effective Mass}
Since the form of \Eq{eqn:f} is designed to capture a band-warped energy dispersion at a critical point in the BZ, we may still use the expression in \Eq{eqn:DOSeffectivemass} to define the DOS effective mass for a warped energy band minimum, or its straightforward modification for an energy band maximum.  Comparing \Eq{eqn:DOS} with \Eq{eqn:DOSeffectivemass}, and recalling the definitions of the numerical factors given in \Eq{eqn:Cdefs}, we may generally define the DOS effective mass as
\begin{equation}\label{eqn:mdosdef}
m_* \equiv \pm \left( \frac{C_{\pm}}{2 \pi} \right)^{2/3} \me.
\end{equation}
%
%
%
%
%
%
\subsection{A Band Warping Parameter}\label{sec:warping}
There are multiple ways of introducing parameters that provide some measures of band warping. However, no single parameter can be expected to account entirely for the full angular complexity of $f(\theta,\phi)$. We have previously introduced one measure of band warping by defining a parameter\cite{MRPF}  
\begin{align}\label{eqn:warping}
w &= \frac{\langle  \, \, (Tr[H] - \langle Tr[H] \rangle )^2 \, \, \rangle^{1/2}}{\langle Tr[H] \rangle}. \nonumber\\
\end{align}

For the sake of illustration, let us return to a two-dimensional energy dispersion as in \Eq{basictwodimensional}, namely, $E = \frac{\hbar^2 k^2}{2 \me} f(\theta) $, and let us further assume that $f(\theta)$ is positive everywhere. Then we have\cite{MRPF} 
\begin{align}\label{eqn:warpingtrace}
\langle \cdot \rangle_{\theta} &= \frac{1}{2 \pi} \int_{0}^{2 \pi} \cdot \, \, \,  \dd \theta, \nonumber\\
Tr[H] &= 2 f(\theta) + 2 f(\theta + \frac{\pi}{2}). 
\end{align}
This definition of the band warping parameter, $w$, essentially measures the coefficient of variation of the sum of the eigenvalues of the Hessian matrix $H(\theta)$ formally obtained in each Cartesian coordinate system rotated by an angle $\theta$. A twice-differentiable, i.e., a non-warped surface cannot have any variation of its quadratic form eigenvalues. Hence, that must have $w=0$. However, the converse is not necessarily true. Namely, having $w=0$ is not sufficient to conclude that the energy dispersion is twice-differentiable. The same considerations also apply to $f(\theta)$ surfaces with positive and negative values in different regions of $S^2$, as indicated in \S \ref{sec:dos}.

\subsection{The DOS Effective Masses for the Kittel Form}
As a basic illustration of our results, let us calculate the DOS effective masses for the heavy-hole (hh) and light-hole (lh) bands described by what we may dub the ``Kittel form," originally derived in a ground-breaking paper,\cite{Dresselhaus} as

%
\begin{align}\label{eqn:kittel}
&E(\bd{k}) = \nonumber\\ &\frac{\hbar^2}{2 \me} \left( A k^2 \pm \left[ B^2 k^4 + C^2 (k_x^2 k_y^2+ k_y^2 k_z^2 + k_z^2 k_x^2) \right]^{1/2} \right).
\end{align}
Expressing that according to our \Eq{eqn:f}, we obtain exactly\cite{MRPF}
\begin{align}\label{eqn:kittelf}
&f(\theta, \phi) = A \pm \\ \nonumber
& \sqrt{B^2+C^2 \sin ^2(\theta ) \left[\cos^2 (\theta) + \cos^2 (\phi) \sin^2 ( \theta ) \sin^2 (\phi) \right]}.
\end{align}
In both expressions, the upper positive (lower negative) sign refers to the heavy (light) hole band dispersion. We use again Rydberg atomic units, where $\hbar^2/ 2 \me = 1$.

\begin{figure}[!hb]
	\begin{center}
 	\includegraphics[width=\figsize]{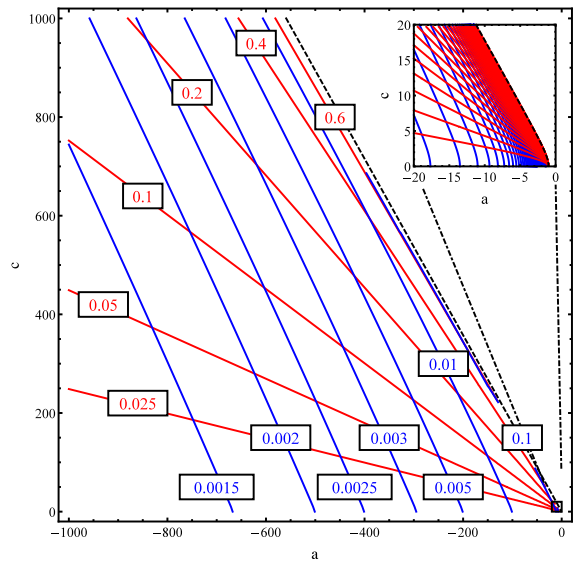}\\
      \caption{\label{fig:fig1}  Contour plots of constant the DOS heavy-hole effective mass $m_{\textrm{hh}}$, in units of $\me B$, are shown as blue curves. For comparison, contour plots of constant absolute value of warping parameter $w$ are shown as red curves.}
	\end{center}
\end{figure}

Although we may not be able to express it in a closed analytic form, each DOS effective mass for the Kittel form can be evaluated numerically using \Eq{eqn:Cdefs} and \Eq{eqn:mdosdef}. Let us further factorize the $B$ parameter in front of the energy dispersion of the Kittel form or its angular effective mass surface. Contour plots of the corresponding DOS heavy-hole effective mass, $m_{\textrm{hh}}$, as functions of $a=A/|B|$ and $c=C/|B|$, are shown in blue in Fig.\ \ref{fig:fig1}. Numerical values of $m_{\textrm{hh}}$ are given in units of $\me B$. Notice that $f(\theta,\phi)$ becomes imaginary for some values of $\theta$ and $\phi$ if $c$ exceeds a $c_{\textrm{max}}$ given by
\begin{equation}\label{eqn:max}
c_{\textrm{max}} (a) = \frac{4 \sqrt{a^2 - 1}}{\sqrt{5}}.
\end{equation}  
Contours of constant $m_{\textrm{hh}}$ thus appear to accumulate along a corresponding curve.  It is not clear whether any $m_{\textrm{hh}}$ may be attained for values of $a$ and $c$ approaching \Eq{eqn:max} from below. 

We may also compute the band warping parameter, $w$, for the heavy-hole band of the Kittel form, based on the analog of \Eq{eqn:warping} to three-dimensional energy dispersions. Contour plots of constant $w$ are shown in red in \Figref{fig:fig1}. Notice that, moving along curves of constant $w$, the DOS heavy-hole effective mass $m_{\textrm{hh}}$ increases with increasing $a$. Alternatively, moving along curves of constant $m_{\textrm{hh}}$, the band warping parameter $w$ increases with increasing $c$. Thus, perhaps surprisingly, a larger value of $w$ does not necessarily imply either a larger or a smaller value of $m_{\textrm{hh}}$, since that depends on the values of $a$ and $c$ parameters; and conversely.

Our results for the angular version of the DOS effective mass are in fact consistent with those of an original paper by Lax and Mavroides,\cite{laxmavroides} if one identifies their $g(\theta,\phi)$ with the precise angular effective mass surface $f(\theta,\phi)$ introduced in Ref.~\onlinecite{MRPF} and used in this context.

\section{Effects of Band Warping on the DOS and the DOS Effective Masses}\label{sec:Kittel}

Given the somewhat unexpected results that we have obtained for the Kittel form, it is natural to question what effects or relations may generally exist between band warping and the DOS effective masses. For example, if we consider energy dispersions with angular contributions giving rise to finite $C_{\pm}$ in \Eq{eqn:Cdefs}, then the only effect that band warping can have on the DOS is to modify that numerical factor in front of the square-root energy dependence in \Eq{eqn:DOS}. Additional insight about the integrated contribution of $f(\theta,\phi)$ in \Eq{eqn:Cdefs} may be gained by using methods similar to that of a stationary phase, that is, by looking for particular directions that may predominantly contribute to the overall DOS effective mass. In any case, it is clear that the DOS can be increased by increasing the effective mass given by \Eq{eqn:mdosdef} in \Eq{eqn:DOS}. 

\begin{figure}[!hb]
	\begin{center}
	\includegraphics[width=\figsize]{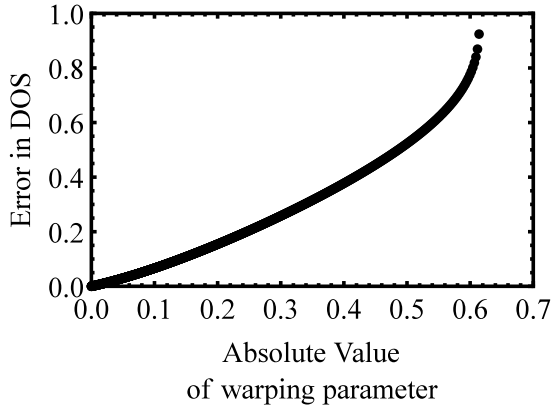}\\
    \caption{\label{fig:fig2} Relative error versus band warping parameter, $w$, for the Kittel form. For specificity, we start at a point $(-\sqrt{312501} \approx -559,0)$ in \Figref{fig:fig1}, and then we increase $c$ vertically. Evidently, the relative error of the DOS effective mass, derived from the least-squares fit of the angular effective mass surface to a spherical surface, increases monotonically with  $w$.}	
\end{center}
\end{figure}

For the Kittel form, our $w$ parameter may also be used to indicate how far from spherical is the angular effective mass surface $f(\theta,\phi)$. For example, in the plane $(a,c)$ of Fig.\ \ref{fig:fig1}, if we climb vertically along the positive $c$ axis from some point, e.g. $(-\sqrt{312501},0)$, $w$ increases.\footnote{The particular $a=-\sqrt{312501}$ value was chosen just to let $c$ range from 0 to 1000.} Let us then compute the error between an approximate DOS effective mass, derived from the least-squares fit of the $f(\theta,\phi)$ surface to a sphere, and the correct DOS effective mass, calculated from \Eqn{eqn:Cdefs}. That error is plotted in \Figref{fig:fig2}. As expected, when $c = 0$, the relative error ($\frac{|value-exact|}{exact}$) is zero, because the angular effective mass surface is actually spherical.  However, as $c$ and $w$ increase, the relative error ($\frac{|value-exact|}{exact}$) increases up to almost 100\%! Thus, at least for the Kittel form, we may say that $w$ provides some indication of how far is the energy dispersion from being twice-differentiable .

\section{Relations to the Lax-Mavroides and Lawaetz DOS Effective Masses}\label{sec:Lawaetzetal}
Lax and Mavroides\cite{laxmavroides} originally proposed the correct idea of an angular effective mass, but they immediately contaminated it with questionable expansions meant to fit the Kittel form specifically. Their Eq.\ (8) and those at the beginning of their Sec.\ IIIA correspond to our Equations (\ref{eqn:Cdefs}) and (\ref{eqn:mdosdef}),
in defining the DOS effective mass. However, not only is our treatment much more general than theirs, but it also applies more appropriately to the Kittel form, based on \Eq{eqn:kittelf}.

\begin{figure}[!hb]
	\begin{center}
 	\includegraphics[width=\figsize]{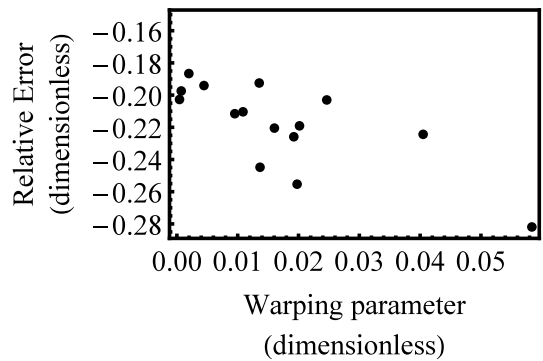}\\
      \caption{\label{fig:fig3} Relative error of the DOS heavy-hole effective masses $m_{\textrm{hd}}$ estimated by Lawaetz\cite{Lawaetz1971} and reported in column 5 of Table I, as compared to our correct values, computed from \Eq{eqn:Cdefs} and \Eq{eqn:mdosdef} and reported in column 6 of Table I, versus the band warping parameter $w$. }
	\end{center}
\end{figure}

Our treatment of the DOS effective masses is also much more rigorous and clearer than that of Lawaetz.\cite{Lawaetz1971} Using our generally correct expressions and integrating them numerically for the same values of parameters reported by Lawaetz for various materials, there are significant differences between our appropriate DOS effective masses and those artificially produced by Lawaetz. We show that in Table\ \ref{tab:table1}, where we have used the following relations between the $A,B,$ and $C$ parameters of the Kittel form and the $\gamma_1, \gamma_2,$ and $\gamma_3$ parameters introduced by Luttinger,\cite{LuttingerPR102-1956}
\begin{align}\label{eqn:LKtoK}
A(\gamma_1) &= - \gamma_1, \nonumber\\
B(\gamma_2) &= 2 \gamma_2, \\
C(\gamma_2, \gamma_3) &= \sqrt{12 (\gamma_3^2 - \gamma_2^2)}.\nonumber
\end{align}
\begingroup
\squeezetable
\begin{table}
\caption{ \label{tab:table1}Comparison of the DOS effective masses for materials reported in Table II of Ref.\ \onlinecite{Lawaetz1971} and those correctly derived from our Equations (\ref{eqn:Cdefs}) and (\ref{eqn:mdosdef}).}
\begin{ruledtabular}
\begin{tabular}{cccccccc}
\hline
Crystal & $\gamma_1$ & $\gamma_2$ & $\gamma_3$ & Lawaetz           & Correct           & Lawaetz            & Correct \\
        &            &            &            & $m_{\textrm{hd}}$ & $m_{\textrm{hd}}$ & $m_{\textrm{ld}}$  & $m_{\textrm{ld}}$\\
\hline
C & 4.62 & -0.38 & 1. & $\phantom{}$\footnote{Formalism invalid because $\gamma_2$ and $\gamma_3$ have opposite sign} & $\phantom{}^{a}$ & $\phantom{}^{a}$ & $\phantom{}^{a}$\\ 
Si & 4.22 & 0.39 & 1.44 & 0.53 & 0.537
& 0.16 & 0.156
\\ 
Ge & 13.35 & 4.25 & 5.69 & 0.35 & 0.351
& 0.043 & 0.0423
\\ 
Sn & -14.97 & -10.61 & -8.52 & 0.29 & 0.289
& -0.029 & -0.0297
\\ 
AlP & 3.47 & 0.06 & 1.15 & 0.63 & 0.615
& 0.2 & 0.195
\\ 
AlAs & 4.04 & 0.78 & 1.57 & 0.76 & 0.752
& 0.15 & 0.151
\\ 
AlSb & 4.15 & 1.01 & 1.75 & 0.94 & 0.953
& 0.14 & 0.141
\\ 
GaP & 4.2 & 0.98 & 1.66 & 0.79 & 0.786
& 0.14 & 0.143
\\ 
GaAs & 7.65 & 2.41 & 3.28 & 0.62 & 0.620
& 0.074 & 0.0739
\\ 
GaSb & 11.8 & 4.03 & 5.26 & 0.49 & 0.498
& 0.046 & 0.0468
\\ 
InP & 6.28 & 2.08 & 2.76 & 0.85 & 0.858
& 0.089 & 0.0887
\\ 
InAs & 19.67 & 8.37 & 9.29 & 0.60 & 0.600
& 0.027 & 0.0267
\\ 
InSb & 35.08 & 15.64 & 16.91 & 0.47 & 0.490
& 0.015 & 0.0147
\\ 
ZnS & 2.54 & 0.75 & 1.09 & 1.76 & 1.796
& 0.23 & 0.224
\\ 
ZnSe & 3.77 & 1.24 & 1.67 & 1.44 & 1.468
& 0.149 & 0.148
\\ 
ZnTe & 3.74 & 1.07 & 1.64 & 1.27 & 1.296
& 0.154 & 0.152
\\ 
CdTe & 5.29 & 1.89 & 2.46 & 1.38 & 1.466
& 0.103 & 0.102
\\ 
HgS & -41.28 & -21 & -20.73 & 2.78 & 2.946
& -0.012 & -0.0121
\\ 
HgSe & -25.96 & -13.69 & -13.2 & 1.36 & 1.341
& -0.019 & -0.0190
\\ 
HgTe & -18.68 & -10.19 & -9.56 & 1.12 & 1.220
& -0.026 & -0.0261
\\
\hline
\end{tabular}
\end{ruledtabular}
\end{table}
\endgroup

Figure \ref{fig:fig3} shows the error of the DOS heavy-hole effective mass estimated by Lawaetz and its correlation with our warping parameter $w$ for that band in various materials. That error is partly the result of inconsistent series expansions and truncations in procedures elaborated by Lax, Mavroides and Lawaetz.\cite{laxmavroides, mavroideslaxPR107in1957, Lawaetz1971} Roughly, the larger is warping or $w$, the greater is the discrepancy between Lawaetz's estimate and our precise determination of the DOS effective mass. That error can be quantitatively as large as 28\%. More importantly, the original lack of a precise definition and treatment of warped bands has been responsible for a lack of consistency among many subsequent papers and \emph{ad hoc} estimates of the DOS effective masses. 
 
To illustrate more subtle effects of band warping on the DOS, we shall further consider some two-dimensional cases where we can quantitatively control parameters that provide different measures of band warping, namely, either the $w$ parameter that we have already introduced, or an alternative band warping parameter to which we may refer more generally as band ``corrugation."

\section{Two-Dimensional Cases}\label{sec:Two-Dimensional}
\subsection{Two-dimensional Kittel form}
\begin{figure}[!hb]
	\begin{center}
	\includegraphics[width=\figsize]{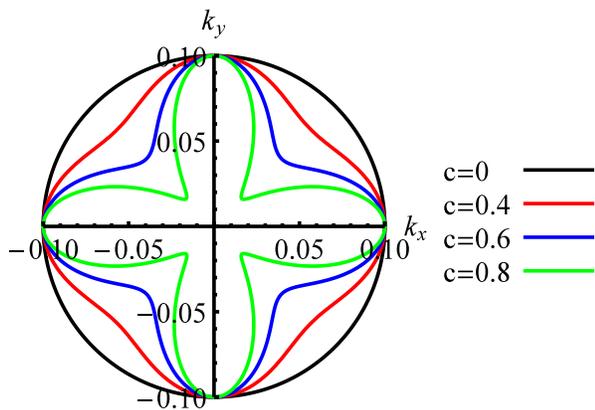}\\
    \caption{\label{fig:fig4} Angular effective mass contours of $f(\theta)$ for the two-dimensional Kittel form that has $k_z = 0$, for parameter values of $a=-1.1$ and $c=0.0, 0.4, 0.6$, and $0.8$. The $k$ dependence is exactly parabolic in every radial direction.}	
\end{center}
\end{figure}
As a first case, consider a two-dimensional version of the Kittel form determined by setting $k_z = 0$ in \Eq{eqn:kittel}, namely,
\begin{align}\label{eqn:ex1}
E(k_x, k_y) &= \frac{\hbar^2}{2 \me} ( A k^2 \pm \sqrt{B^2 k^4 + C^2 k_x^2 k_y^2} ) \nonumber \\
&= |B| \frac{\hbar^2}{2 \me} ( a k^2 \pm \sqrt{k^4 + c^2 k_x^2 k_y^2} ).
\end{align}
Equivalently, by setting $\theta = \pi/2$ in \Eq{eqn:kittelf}, and then relabeling the azimuthal angle $\phi$ with the two-dimensional polar angle $\theta$, we obtain exactly
\begin{align}\label{eqn:ex1a}
E(k, \theta) &=\frac{\hbar^2 k^2}{2 \me} f(\theta)\nonumber\\ 
&= |B| \frac{\hbar^2 k^2}{2 \me} (a \pm \sqrt{1 + c^2 \cos^2 \theta \sin^2 \theta}) .
\end{align}
Angular effective mass planar contours of $f(\theta)$ are shown in \Figref{fig:fig4} for a given value of $a$ and four increasing values of the $c$ parameter. 

In this two-dimensional case,  the band warping parameter, $w$, and the DOS effective mass, $m_*$, can be expressed analytically, for any $c < c_{\textrm{max}}$, as
\begin{widetext}
\begin{subequations}\label{eqn:wandm}
\begin{align}
w &= \frac{\sqrt{2} \sqrt{\pi ^2 \left(c^2+8\right)-8 E\left(-\frac{c^2}{4}\right)^2-8 \sqrt{c^2+4} E\left(\frac{c^2}{c^2+4}\right) E\left(-\frac{c^2}{4}\right)-2 \left(c^2+4\right) E\left(\frac{c^2}{c^2+4}\right)^2}}{4 \pi  a+4 E\left(-\frac{c^2}{4}\right)+2 \sqrt{c^2+4} E\left(\frac{c^2}{c^2+4}\right)}, \\
m_* &= -\frac{1}{\pi |B|} \frac{2 \left(\sqrt{\left(a^2-1\right) \left(4 a^2-c^2-4\right)} \left(\left(1-a^2\right) K\left(-\frac{c^2}{4}\right)+a^2 \Pi \left(\frac{c^2}{4 \left(a^2-1\right)} \bigg|-\frac{c^2}{4}\right)\right)+\pi a \left(a^2-1\right)\right)}{\left(a^2-1\right)^{3/2} \sqrt{4 a^2-c^2-4}.}
\end{align}
\end{subequations}
\end{widetext}
In \Eq{eqn:wandm}, $E(m)$, $K(m)$, and $\Pi(n,m)$ denote the complete elliptic integral, the complete elliptic integral of the first kind, and the complete elliptic integral of the third kind, respectively, and $m = \sin^2 \alpha$ and $n$ are their standard arguments. 

Contours of constant DOS heavy-hole effective mass $m_{\textrm{hh}}$ and contours of constant absolute value of warping parameter $w$ for this two-dimensional Kittel form are qualitatively similar to those of the full three-dimensional Kittel form, which was shown in \Figref{fig:fig1}. In two dimensions $w$ attains a maximum magnitude whenever $a$ and $c$ approach the limit of $c_{\textrm{max}}$. In two dimensions, that is
\begin{equation}
w_{\textrm{max}} = -\frac{\sqrt{\frac{1}{2} \left(\pi ^2-8\right)}}{\pi -2} \approx -0.8447. 
\end{equation}
We did not investigate a corresponding effect in the full Kittel form but we expect similar results. 

\subsection{Example of $a (1 + b \cos 4 \theta$ )}
\begin{figure}[!hb]
	\begin{center}
	\includegraphics[width=\figsize]{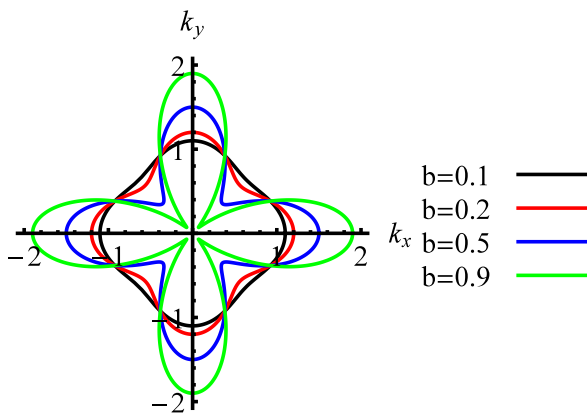}\\
    \caption{\label{fig:fig5} Angular effective mass contours of $f(\theta)$ for a two-dimensional dispersion relation of the form $E = \frac{\hbar^2 k^2}{2 \me} (a + b \cos 4 \theta)$, where we set $a=1$, and $b=0.1, 0.2, 0.5$, and $0.9$.}	
\end{center}
\end{figure}

As a second example, we consider the two-dimensional energy dispersion
\begin{equation}\label{eqn:ex2}
E = \frac{\hbar^2 k^2}{2 \me} f(\theta) = \frac{\hbar^2 k^2}{2 \me} a (1 + b \cos 4 \theta).
\end{equation}
Unless $b=0$, this function is not twice-differentiable at the origin exclusively, as an isolated point. In Fig.~\ref{fig:fig5}, its angular effective mass $f(\theta)$ is plotted for $a=1$ and four increasing values of $b$. Since the integral in \Eq{eqn:Cdefs2D} involves $|f^{-1}(\theta)|$, we can indefinitely decrease $m_*$ in \Eq{eqn:mdosdef} by letting $a$ become as small as we need. On the other hand, for any given value of $a$, we expect substantial contributions to $m_*$ from diagonal directions, along which $|f(\theta)|$ becomes increasingly smaller with increasing $b$ values approaching $1^{-}$. In fact, analytic derivations yield
\begin{align}
w &= \frac{b}{\sqrt{2}} , \\
m_* &= \frac{1}{a \sqrt{1-b^2}}, 
\end{align}
where $m_* = C_+/\pi$.
%
%
%
%
So, the band warping parameter, $w$, and the DOS effective mass, $m_*$, are independent of each other, since only $m_*$ depends on $a$, as expected. This behavior may have not been anticipated, but of course we constructed this illustration for that purpose.

\subsection{Example of $n^2 \left( \cos^{2 n} n \theta + \sin^{2 n} n \theta \right)$ }
\begin{figure}[!hb]
	\begin{center}
 	\includegraphics[width=\figsize]{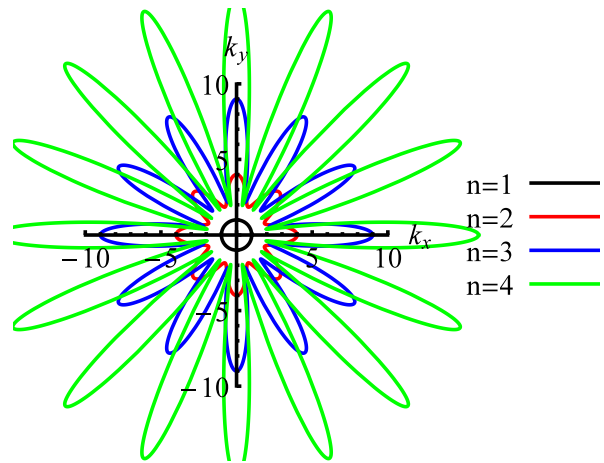}\\
      \caption{\label{fig:fig6} Angular effective mass contours of $f(\theta)$ for a two-dimensional dispersion relation of the form $E = \frac{\hbar^2 k^2}{2 \me} n^2 (\cos^{2 n} n \theta+ \sin^{2 n} n \theta)$ for $n = 1$, 2, 3, and 4.}
\end{center}
\end{figure}
Let us now provide a more complex example where $w$ steadily increases with what we may dub band ``corrugation," whereas $m_*$ at first decreases, but then increases with that ``corrugation." Consider an energy dispersion of the form
\begin{equation}\label{eqn:ex3}
E = \frac{\hbar^2 k^2}{2 \me} f(\theta) =  \frac{\hbar^2 k^2}{2 \me} (n^2 \left( \cos^{2 n} n \theta+ \sin^{2 n} n \theta \right) ).
\end{equation}
Again, unless $n=1$, this function is not twice-differentiable at the origin exclusively, as an isolated point. In \Figref{fig:fig6} we show plots of its angular effective mass $f(\theta)$ for $n = 1$, 2, 3, and 4.

The impression conveyed by \Figref{fig:fig6} is that the energy dispersion ought to deviate more and more from being twice-differentiable with increasing $n$. We are thus led to regard $n$ as a separate parameter, independent of $w$, that may provide an alternative, albeit qualitative, measure of band warping. So, we associate with $n$ a name and a notion of band ``corrugation," although that can hardly provide or lead to any more rigorous or general definition. In any case, the basic idea of ``corrugation" is that it increases with increasing number of radial ``valleys." One may thus expect that the DOS effective mass also increases correspondingly. Although that is often the case, it may not always be so, as we demonstrate with this example. In fact, we could provide many more examples where the DOS effective mass increase or decrease with, or remains independent of, ``corrugation.''

\begin{figure}[!hb]
	\begin{center}
 	\includegraphics[width=\figsize]{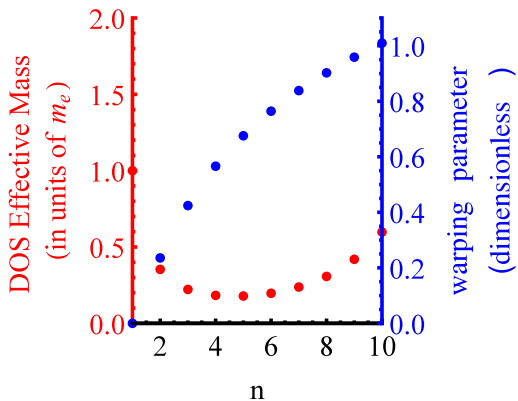}\\
      \caption{\label{fig:fig7} Warping parameter $w$ (in blue) and the DOS effective mass $m_*$ (in red and in units of $\me$) corresponding to \Eq{eqn:ex3}. While $w$ increases monotonically with $n$, $m_*$ decreases at first, but subsequently increases with $n$.}
	\end{center}
\end{figure}

In this example, we can still derive analytic expressions for the band warping parameter, $w$, and the DOS effective mass, $m_*$. However, those expressions are fairly elaborate and we omit them for the sake of conciseness. Suffice it to say that $w$ increases monotonically with $n$, whereas $m_*$ at first decreases with $n$, but then it reaches a minimum, after which $m_*$ increases monotonically with $n$. Corresponding plots of $w$ and $m_*$ are shown in \Figref{fig:fig7}. This example thus demonstrates that $w$ and $m_*$ do not necessarily correlate with each other, nor with the notion of band ``corrugation." 

\subsection{Corrugated example with $w=0$}
We have already demonstrated that the warping parameter $w$ may not necessarily increase or correlate with an increasing DOS effective mass $m_*$. In fact, looking back at \Figref{fig:fig1}, we can easily draw parametrized curves where $w$ decreases while $m_*$ increases. We can also draw curves in \Figref{fig:fig1} where $w$ stays constant while $m_*$ either increases or decreases.  

\begin{figure}[!hb]
	\begin{center}
 	\includegraphics[width=\figsize]{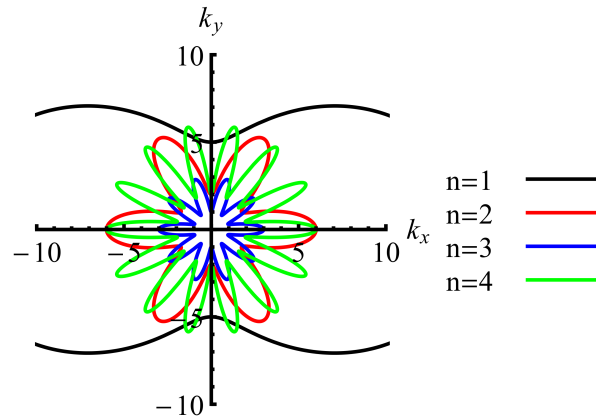}\\
      \caption{\label{fig:fig8} Angular effective mass contours of $f(\theta)$ for a two-dimensional dispersion relation of the form $E = \frac{\hbar^2 k^2}{2 \me} f(\theta) =  \frac{\hbar^2 k^2}{2 \me} ((n^2 - 10 n + 30) \left( 2 + \cos 2 (2 n -1) \theta \right) )$ for $n = 1$, 2, 3, and 4.}
\end{center}
\end{figure}

Let us then provide a conclusive two-dimensional example that has $w=0$, although the energy dispersion is \textit{not} twice-differentiable, and $m_*$ still decreases at first, and then increases with increasing corrugation or $n$. Consider the energy dispersion $E = \frac{\hbar^2 k^2}{2 \me} f(\theta) =  \frac{\hbar^2 k^2}{2 \me} ((n^2 - 10 n + 30) \left( 2 + \cos 2 (2 n -1) \theta \right) )$, whose angular effective mass $f(\theta)$ is plotted in \Figref{fig:fig8} for $n = 1$, 2, 3, and 4. We can prove that the band warping parameter is $w=0$, independently of corrugation or $n$, but the DOS effective mass $m_*$ at first increases with corrugation, then it reaches a maximum at $n=3$, and subsequently decreases monotonically for all $n>3$.

This example demonstrates in particular that a function that is \textit{not} twice-differentiable can still have $w=0$. This prompts us to introduce in an Appendix a more refined definition of a band warping parameter $\mathscr{W}$ that captures at least that type of non-differentiability. The $\mathscr{W}$ alternative to $w$ is more elaborate but possibly more helpful in identifying energy dispersions that are not twice-differentiable. However, since second-order differentiability is inherently based on multi-dimensional limits, there can be no single parameter whose vanishing is sufficient to guarantee that any particular energy dispersion is certainly twice-differentiable at a point.

\section{Conclusions}\label{sec:Conclusions}
We have applied the angular effective mass formalism introduced in Ref.\ \onlinecite{MRPF} to study the density of states in warped and non-warped energy bands at critical points in the Brillouin zone. First we have verified ordinary results for ellipsoidal and hyperbolic energy dispersions. Then we have generalized the expression of the DOS to account for general band warping and monotonically increasing non-parabolic energy dispersions. Band warping may or may not increase the DOS effective mass. An intuitive notion of greater band ``corrugation," referring to energy dispersions that deviate ``more severely" from being twice-differentiable at an isolated critical point, may also vary independently of the corresponding DOS effective mass and band warping parameter. We have demonstrated these effects through investigation of valence band energy dispersions in cubic materials, showing the superiority of the angular effective mass treatment of the DOS effective masses compared to that of original papers.\cite{Dresselhaus, laxmavroides, mavroideslaxPR107in1957, Lawaetz1971}  

We have further considered certain two-dimensional physical and mathematical examples that may be relevant to studies of band warping in heterostructures\cite{shechter1995orientation, ChenFornariResca, simion2014magnetic} and surfaces\cite{goldoni1991band}. These examples may also be useful in clarifying the interplay between possible band warping and band non-parabolicity for non-degenerate conduction band minima in thermoelectric materials of corresponding interest.\cite{Chen2013, ParkerPRL2013, Singh2015May} 

\acknowledgments
This work was supported by the Vitreous State Laboratory of The Catholic University of America. MF acknowledges collaboration with the AFLOW Consortium (http://www.aflowlib.org) under the sponsorship of DOD-ONR (N000141310635).

\section{Appendix}\label{sec:Appendix}

If $f(x,y)$ is a twice-differentiable function of two Cartesian variables at a point $\bd{a}$, then the second-order directional derivative at $\bd{a}$ is defined as
\begin{widetext}
\begin{equation}
D_{\theta}^2 f(a_x,a_y) = \lim_{t \to 0} \frac{f(a_x + 2 t \cos(\theta), a_y + 2 t \sin(\theta) ) - 2 f(a_x + t \cos(\theta), a_y + t \sin(\theta)) + f(a_x,a_y)}{t^2}.
\end{equation}
\end{widetext}
This directional derivative can also be expressed as a linear combinations of second-order partial derivatives along the coordinate $x$- and $y$- axes at $\bd{a}$. 

If $\mathbb{A}$ is an orthogonal matrix with determinant +1, whose first column is derived from the $x$-axis rotated into a new direction by an angle $\theta$, and $\bd{\hat{x}}$ is the unit vector along the original $x$-axis, then one can show that
\begin{equation}\label{eqn:directional}
D_{\theta}^2 f(a_x,a_y) = \bd{\hat{x}}^{\intercal} \mathbb{A}^{\intercal} H(\theta) f(\bd{a}) \mathbb{A} \bd{\hat{x}},
\end{equation}
where $H(\theta)$ is the Hessian matrix of ordinary second-order partial derivatives at $\bd{a}$.

If $f(x,y)$ is twice-differentiable at $\bd{a}$, the band warping parameter that we have previously introduced must vanish.\cite{MRPF} Namely, $w=0$ is a necessary condition for second-order differentiability at a critical point. However, $w=0$ is \textit{not} a sufficient condition for second-order differentiability at a critical point. Expecting that any single parameter could capture the full complexity of $f(x,y)$ around $\bd{a}$ would indeed be asking too much.

For example, consider $g(x,y)$ defined as zero at the origin and $g(x,y) = \dfrac{3 x^6-9 x^4 y^2+21 x^2 y^4+y^6}{\left(x^2+y^2\right)^2}$ everywhere else. That has $w=0$, although $g(x,y)$ is \textit{not} twice-differentiable at the origin, exclusively. In fact, its angular function is $(2 +  \cos(6 \theta)) r^2$, which corresponds to the last example given in the previous text for $n=1$.

More generally, any function of the form $r^2 f(\theta)$, where 
\begin{equation}
f(\theta) = a_0 + \sum_{n=1}^{\infty} a_n \cos( 2 (2 n-1) \theta) ,\\
\end{equation}
shares the same peculiarity.  

These considerations prompt us to consider other measures of band warping for functions that are not twice-differentiable. Consider, for example, the difference between the second-order directional derivative at $\bd{a}$ and the correspondingly rotated Hessian matrix element, namely,
\begin{equation}
\delta (\theta) = D_{\theta}^2 f(a_x,a_y) - \bd{\hat{x}}^{\intercal} \mathbb{A}^{\intercal} H(\theta) f(\bd{a}) \mathbb{A} \bd{\hat{x}}.
\end{equation}
According to \Eq{eqn:directional}, if $f(x,y)$ is twice-differentiable at $\bd{a}$, then $\delta (\theta)$ must vanish for any $\theta$. Unfortunately, the converse still cannot provide a sufficient condition in general. Nevertheless, detecting a non-vanishing $\delta (\theta)$ at any $\theta$ may provide a more refined tool to discover whether $f(x,y)$ is \textit{not} twice-differentiable. That is accomplished by evaluating the single parameter 
\begin{equation}\label{eqn:superw}
\mathscr{W} = \left( \frac{1}{2 \pi} \int_{0}^{2 \pi} \left( \delta(\theta) \right)^2 \, \,  \dd \theta \right)^{1/2} .
\end{equation}
For the previous example of $g(x,y)$ with $w=0$, \Eq{eqn:superw} indeed provides a non-vanishing $\mathscr{W} = 34$.  Thus, if we use $\mathscr{W}$ rather than $w$, we can conclude that $g(x,y)$ is \textit{not} twice-differentiable at the origin. Most generally, however, even $\mathscr{W} = 0$ cannot guarantee that any particular function $f(x,y)$ is certainly twice-differentiable at a point.

\end{document}